\documentstyle[aps,twocolumn,prl,tighten,epsfig]{revtex} 
\input epsf
\def\plotone#1{\centering \leavevmode
\epsfxsize= 0.95\columnwidth \epsfbox{#1}}

\def\MeV{\,{\rm MeV}}

\def\rcm{\,{\rm cm}}

\def\ev{{\,\rm eV}}

\def\cmm2{{\,\rm cm^{-2}}}
\def\cm2{{\,{\rm cm}^2}}
\def\cmm3{{\,{\rm cm}^{-3}}}
\def\gcmm3{{\,{\rm g\,cm^{-3}}}}

\def\fun#1#2{\lower3.6pt\vbox{\baselineskip0pt\lineskip.9pt
  \ialign{$\mathsurround=0pt#1\hfil##\hfil$\crcr#2\crcr\sim\crcr}}}

\def\prd{Phys. Rev. D~}
\def\prl{Phys. Rev. Lett.~}
\def\apj{Astrophys. J.~}
\def\apjl{Astrophys. J. Lett.~}

\def\eg{{\it e.g.}}
\def\ie{{\it i.e.}}
\def\etal{{\it et al.}}
\def\apriori{{\it a priori}}

\def\evcmm3{\ev\rcm^{-3}}

\def\zls{z_{\rm LS}}

\def\nnu{N_\nu^{\rm equiv}}

\begin{document}
\twocolumn[\hsize\textwidth\columnwidth\hsize\csname @twocolumnfalse\endcsname
\draft

\title{Precision Cosmology and the Density of Baryons in the Universe}

\author{Manoj\ Kaplinghat$^{1}$ and Michael\ S.\ Turner$^{1,2,3}$}
\address{$^1$ Department of Astronomy and Astrophysics\\
The University of Chicago, 5640 S. Ellis Ave., Chicago, IL 60637, USA}
\address{$^2$ NASA/Fermilab Astrophysics Center\\
Fermi National Accelerator Laboratory, PO Box 500\\
Batavia, IL  60510-0500 USA}
\address{$^3$ Department of Physics, Enrico Fermi Institute\\
The University of Chicago, Chicago, Illinois 60637 USA}

\maketitle

\begin{abstract}
Big-bang Nucleosynthesis (BBN) and Cosmic Microwave Background
(CMB) anisotropy measurements give independent, accurate measurements
of the baryon density and can test the framework of the standard
cosmology.  Early CMB data are consistent with the longstanding
conclusion from BBN that baryons constitute a small fraction of
matter in the Universe, but may indicate a slightly higher value for the
baryon density. We clarify precisely what the two methods 
determine, and point out that differing values for the baryon density 
can indicate either an inconsistency or physics beyond the standard
models of cosmology and particle physics. We discuss other signatures of
the new physics in CMB anisotropy.
\end{abstract}
]

{\parindent0pt\it Introduction.}
Just a decade ago the phrase ``precision cosmology''
would have been an oxymoron.  The COBE
FIRAS determination of the temperature of the Cosmic Microwave
Background (CMB) to four significant figures, 
$T_0 = 2.725\pm 0.001 \,{\rm K}$
\cite{mather99} (we quote all errors at $1\sigma$)
should dispel such thoughts.  Cosmologists now foresee a
precision era where a flood of high-quality data, from measurements
of CMB anisotropy to large-scale structure, pin down
cosmological parameters to percent-level precision,
decisively testing theories of the early Universe and
probing physics at energy scales beyond those accessible in
accelerator experiments \cite{tyson}.

Some of the data can also test the consistency
of the standard cosmology.  In particular, 
the longstanding big-bang nucleosynthesis determination
of the baryon density, will be checked by CMB anisotropy data at
one percent or better \cite{dns_mst_99}.  The physics
underlying the two measurements could hardly be more different:
the baryon density at
a time of one second determines how complete the conversion of
neutrons and protons to tightly-bound $^4$He nuclei is, while the
baryon density 400,000 years later (at the time of last scattering)
determines the amplitude of gravity-driven acoustic oscillations
in the baryon-photon fluid.

The determination of the primeval deuterium abundance in a number of
high-redshift ($z\sim 2-4$) hydrogen clouds
\cite{tytler_00} coupled with refined predictions
of the standard theory of BBN has led to a determination of the
baryon density to an accuracy of about 5\%,
$\Omega_B h^2 = 0.019\pm 0.00095$ \cite{bntt}.
Very recently, the BOOMERanG \cite{boom00} and MAXIMA \cite{max00}
balloon-borne CMB anisotropy experiments have mapped CMB anisotropy
with sufficient angular resolution to make the first
CMB determinations of the baryon density, $\Omega_Bh^2 =
0.032^{+0.005}_{-0.004}$ \cite{boom_max}.  While the CMB result
has a much larger uncertainty (and depends upon the 
imposed priors and parameters that are allowed to 
vary \cite{other}), additional data and new CMB measurements should 
soon narrow the gap between the two methods.

The early results are encouraging for the consistency of
the standard cosmology:  the CMB value for the baryon density
agrees with the BBN determination within about $2\sigma$ and lies far from
dynamical determinations of the total mass density, 
$\Omega_M h^2 = 0.2\pm 0.04$
(inferred from $\Omega_M = 0.35 \pm 0.07$ \cite{mst_99} and 
$h=0.7\pm 0.07$ \cite{H0}).
Thus, the CMB measurement strongly supports the case for nonbaryonic dark
matter, whose linchpin for twenty years has been the discrepancy between the
BBN value for $\Omega_B$ and dynamical measurements of $\Omega_M$
\cite{dns_mst_99,os_review}.

The purpose of our {\em Letter} is to clarify exactly what is determined,
with regard to the baryon density, by BBN and CMB anisotropy and to point 
out that the two determinations need not agree to be consistent. A
discrepancy could very well be the signal of new physics. Here and
throughout $\Omega_B$ denotes the fraction of critical density in baryons
today and $h = H_0/100\,{\rm km\,s^{-1}\,Mpc^{-1}}$.
The physical baryon density today, 
$\rho_B = 1.88\times 10^{-29}(\Omega_Bh^2)\,{\rm g/cm^{3}}$.

{\em Comparing apples to apples.}  In the standard theory of BBN
(i.e., isotropic and homogeneous Universe with only the known
particle species and the assumption that all three neutrino species
are light with mass $\ll 1\MeV$, and negligible chemical potentials),
the yields of BBN depend only upon the baryon-to-photon ratio
$\eta$, the neutron mean lifetime and eleven key nuclear cross sections
\cite{dns_mst_99,os_review}.  The uncertainties due to the neutron mean lifetime
and nuclear data have recently been re-evaluated and significantly
reduced \cite{bntt}.   Based upon these predictions and uncertainties,
the Burles-Tytler primeval deuterium measurement,
${\rm (D/H)}_P = (3.4\pm 0.3)\times 10^{-5}$, implies a baryon-to-photon ratio
$\eta_{\rm BBN} = (5.1\pm 0.2)\times 10^{-10}$ \cite{eta_BBN_note}.

In order to infer the present density of baryons one must convert
$\eta_{\rm BBN}$ to a baryon density at the time of BBN by multiplying
by the photon number density,
$n_\gamma({\rm BBN}) = 2\zeta(3)T_{\rm BBN}^3/\pi^2$, and the
mean mass per baryon ($\equiv \bar m$) and then reduce that
density by the volume increase of the Universe since,
\begin{equation}
\rho_B ({\rm today}) = 2\zeta(3) {\bar m}\eta_{\rm BBN}\, 
R_{\rm BBN}^3 \,T_{\rm BBN}^3/\pi^2 R_0^3\,,
\end{equation}
where $R(t)$ is the cosmic scale factor and baryon-number 
conservation since BBN has been explicitly assumed.

Because we do not know the value of scale factor at the time of
BBN \apriori, we cannot proceed without further assumptions.
The standard assumption is adiabaticity:
the constancy of the electromagnetic entropy per unit comoving volume,
which is proportional to $(RT)^3$ \cite{kt}, since BBN.
Allowing for the possibility that the entropy per unit comoving
volume has changed it follows that
\begin{equation}
\rho_B({\rm today}) = {S_{\rm EM}({\rm BBN}) \over S_{\rm EM}({\rm today})}
\,\eta_{\rm BBN} \,{\bar m}\,n_\gamma ({\rm today})\,.
\end{equation}
The number density of photons today is related to the present temperature
of the CMB, $n_\gamma = (410.5\pm 0.5) \,\rcm^{-3}$; dividing by the
critical density we arrive at the key equation,
\begin{equation}\label{omegab_bbn}
\Omega_Bh^2  =  (0.019\pm 0.00095)\,
{S_{\rm EM}({\rm BBN})\over S_{\rm EM}({\rm today})}\,.
\end{equation}
Entropy production after BBN (e.g., by the out-of-equilibrium decay
of a massive particle) can increase $S_{\rm EM}$, which would diminish
the BBN prediction for the baryon density.  It is also possible to
reduce $S_{\rm EM}$ (at the expense of more exotic physics \cite{bh}),
increasing the BBN prediction for the present baryon density.

\begin{figure}[thbp]
\plotone{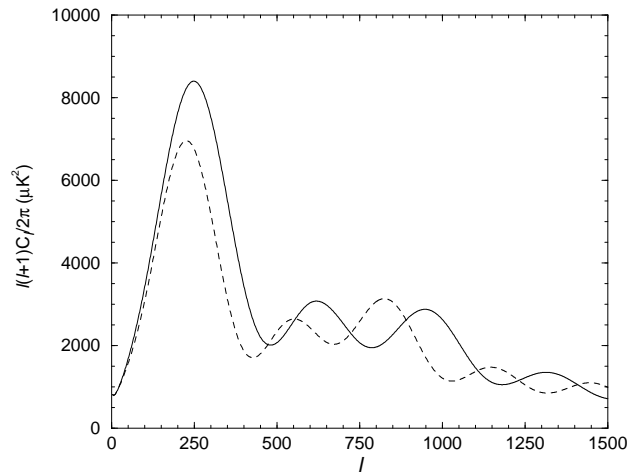}
\caption[caption]{
The multipole power spectrum of CMB anisotropy for
$\nnu = 3,\, 8.3$ (dashed and solid curves respectively) and 
$\Omega_Bh^2 = 0.03$, $\Omega_Mh^2 = 0.17$, $\Omega_0 =1$ and 
$\Omega_\Lambda = 1-\Omega_M$.  Note the characteristic series 
of acoustic peaks and the large change in the power spectrum 
due to the change in $\nnu$.
}
\label{nnu_effect}
\end{figure}

The signature of the baryon density in the CMB involves the
heights of the ``acoustic peaks'' in the angular power spectrum.
Around the time of last scattering the photon -- baryon fluid
is undergoing gravity driven acoustic oscillations; Fourier modes 
caught at maximum compression (odd peaks) or maximum rarefaction 
(even peaks) produce the highest amplitude temperature fluctuations 
on the sky, leading to a series of acoustic peaks in the angular 
power spectrum (see Fig.~\ref{nnu_effect}) \cite{Hu_review}.

The ratio of the heights of the odd and even peaks 
increases with baryon density;
all other cosmological parameters tend to move
the heights of the peaks in unison.  Thus, determining the
baryon density does not suffer
from the cosmic degeneracies that affect other parameters,
and an ultimate precision of better than one percent can be
expected \cite{Hu_review}.

Converting the baryon density
at last scattering (redshift $\zls\simeq 1100$) to the present involves
only a factor of $(1+\zls)^3$ \cite{LS_note}.
While a number of analyses of the BOOMERanG and MAXIMA data have
been carried out \cite{other}, the current state of affairs is
probably fairly represented by the joint analysis of the two 
teams \cite{boom_max},
\begin{equation}\label{omegab_cmb}
\Omega_Bh^2 = 0.032^{+0.005}_{-0.004}\,.
\end{equation}

{\em Changing entropy and its CMB signature.} The entropy increase
due to the out-of-equilibrium decay (\ie, when $T \ll m$) of a 
massive particle relic is given in terms of the particle's 
lifetime and mass \cite{kt}:
\begin{equation}
r \equiv S_f/S_i \approx 3 \,m_{\rm Pl}^{-1/2}\,Y\, m\, \tau^{1/2}\,,
\end{equation}
where $Y\equiv n/s$ is the pre-decay abundance (number density per
unit comoving entropy density -- for a neutrino species, $Y=0.04$).
We have assumed that the decay products thermalize into photons
and increase the EM entropy; to avoid distorting the nearly
perfect black-body spectrum of the CMB this must occur before
about $10^6\,$sec.  In this case the baryon density today is 
smaller (than $\Omega_Bh^2 = 0.019$) by a factor of $r$.

\begin{figure}[thbp]
\plotone{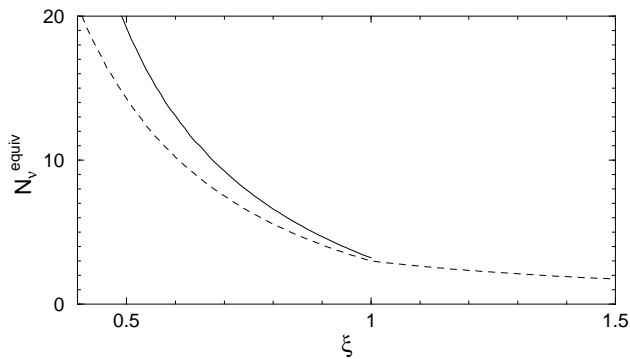}
\caption[caption]{
The relation between $\nnu$ and $\xi$ ($\equiv 0.019/\Omega_Bh^2$) for
the non-standard BBN (solid) and for post-BBN entropy change (dashed). 
Note that for the non-standard BBN case $\xi < 1$ and the plotted
value of $\nnu$ is the minimum required; for the post-BBN entropy
change the plotted curve is the maximum possible.}
\label{r_vs_nnu}
\end{figure}

The increase in entropy has a CMB signature which
involves the fact that the energy density in relativistic particles
at the time of last scattering is smaller than in the standard scenario.
This is because EM entropy production increases the photon-to-neutrino
temperature ratio by a factor of $r^{1/3}$ over the standard value,
$T_\gamma / T_\nu = r^{1/3}(11/4)^{1/3}$, and thereby
decreases the energy in neutrinos at last scattering (which occurs
at a fixed photon temperature, $T_{LS}\simeq 0.3\ev$).  This decrease
in relativistic energy density can be expressed in terms of
a (lower) equivalent number of standard neutrino species
(see Fig. \ref{r_vs_nnu}):
\begin{equation}
\nnu = 3/r^{4/3}\,.
\label{eq:neff1}
\end{equation}
More or less energy in relativistic particles at the time of
last scattering affects the spectrum of CMB anisotropy by
changing the expansion rate and the rate at which the gravitational
potentials associated with density perturbations decay.
Less relativistic energy increases the sound horizon at last
scattering, thereby shifting the acoustic peaks to larger angular
scales (smaller $\ell$).
Less relativistic energy also depresses the power around the 
first acoustic peak by diminishing the integrated 
Sachs-Wolfe effect \cite{Hu_review}, and decreases the damping 
of the higher-order acoustic peaks (because the peaks have shifted
to larger scales relative to the damping length) as can be seen 
in Fig. \ref{nnu_effect}.

Two points deserve mention here: (1) smaller $\nnu$ also changes the
power spectrum of matter inhomogeneity by shifting the epoch of matter --
radiation equality to an earlier time. (2) Only a change in the EM
entropy affects the predicted BBN baryon density; decays without EM decay 
products (\eg, decay to three neutrinos) do not affect the
BBN prediction. If the massive-particle decays produce both EM and
non-EM entropy, the EM decay products determine $r$, and the non-EM decay
products increase the energy density in relativistic particles and lead
to an additional term in Eq.~(\ref{eq:neff1}).

The value of $\nnu$ will ultimately be determined from 
CMB anisotropy to a few percent \cite{lopezetal}.  This provides a 
cross check on this explanation: if the present baryon density inferred
from BBN exceeds that inferred from CMB by a factor
of $\xi$, then $\nnu$ should be $3/\xi^{4/3}$. Of course, the current
data is inconsistent with any post-BBN entropy production, and seems to
require $r < 1$, \ie, entropy reduction.

{\parindent0pt\it Entropy reduction.}  By invoking more exotic physics it
is also possible to reduce the EM entropy.  In order to do so without
violating the 2nd law of thermodynamics, photons must come into thermal
contact with a cooler reservoir (denoted as the X sector) after
BBN and before last scattering.  Such an idea was previous discussed
in a different context \cite{bh}, and we will briefly discuss its possible
relevance here.

The basic idea is simple; owing to the energy dependence of the
cross section for X-sector particles to interact with photons
(and other familiar particles),
the X sector is decoupled (interaction rate per particle $\Gamma$
less than the expansion rate $H$) at high temperatures.  (This can
be achieved provided that $\langle \sigma v \rangle \propto 1/T^n$
with $n>1$.)  At a temperature $T_c$, between the epoch of BBN
and the present, the interactions become rapid enough to
quickly establish thermal contact and $T_X = T$, thereby draining
entropy from the photons (assuming that the X sector was cooler). 

The authors of Ref.~\cite{bh} have shown that a self-consistent model 
for the X sector can be constructed, and further, that it is consistent 
with astrophysical considerations that constrain
the interaction of unseen particles with photons (e.g., emission of
X sector particles through plasmon processes in red giants and supernovae).
With that in mind, let us proceed.

For simplicity, assume that the transfer of energy to X particles
proceeds quickly, by thermally populating massless degrees of freedom
in the X sector with total statistical weight $g_X$.  
The decrease in the EM entropy,
$r=S_{\rm EM}({\rm today})/S_{\rm EM}({\rm BBN})$, follows from energy 
conservation:
\begin{equation}
r = [2/(2+g_X)]^{3/4}\,.
\end{equation}
The present {\em apparent} discrepancy between the BBN and CMB
determinations could be explained with $r\simeq 2/3$ 
($\Rightarrow g_X\simeq 1.5$).

Provided $T_c \geq T_{LS}$, this scenario also leads to an
increase in energy density in
relativistic particles at the time of last scattering and
a signature in CMB anisotropy as discussed above.
The energy increase arises from:  (1) the energy density in X-sector 
particles; and (2) the higher neutrino-to-photon temperature ratio,
$T_\nu /T = r^{-1/3}(4/11)^{1/3}$.  Again, parameterizing this
by the equivalent number of standard neutrino species, it follows that
\begin{equation}\label{r<1}
\nnu = {4\over 7}\left({4\over 11}\right)^{4/3} g_X +3/r^{4/3}\,,
\end{equation}
where the two terms correspond to the two effects just mentioned.

We note that the second term is mandatory and robust -- depressing
the photon temperature necessarily raises the ratio of the neutrino
to photon temperature -- and is identical in form to the term that
arises in the previous case where the entropy is increased,
cf. Eq.~(\ref{eq:neff1}). The first term is model dependent and would
be different if the X sector did not reach the same temperature as
the photons or if only the massive degrees of freedom were excited.
Finally, for $r=2/3$, $\nnu=8.3$, which as Fig.~\ref{nnu_effect} shows has a
dramatic effect on the spectrum of CMB anisotropy.

{\em Concluding remarks.}  By means of very different physics
fine-scale CMB anisotropy and BBN
each have the potential to determine the mean baryon density
to percent accuracy or better.  Currently, the BBN determination
has a precision of 5\% and the CMB measurement 15\%,
and the two disagree at about the $2\sigma$ level.

As we have emphasized, a disagreement between the two determinations
of the baryon density need not indicate inconsistency:
if the BBN baryon density is larger (smaller) by some factor, this 
could be explained by an increase (decrease) in the EM entropy since 
the time of BBN by the same factor \cite{post_bbn_nuc}.
If the two baryon densities agree, then one can limit any post
BBN electromagnetic entropy change.  
As discussed, any entropy change also has a distinctive testable 
signature in CMB anisotropy (see Fig. \ref{r_vs_nnu}).

We have assumed the standard theory of BBN;
relaxing one of its assumptions (e.g., a decaying tau
neutrino with a mass of ${\cal O}(\MeV)$ \cite{tau_decay} or large
neutrino chemical potentials \cite{steigman})
can also change the baryon density inferred
from the primeval deuterium abundance.  Both possibilities have
been discussed \cite{resolve}.  These solutions work by speeding
up the expansion around the end of BBN through additional energy
density in relativistic particles so that more deuterium
remains unburnt.  Thus, both solutions also lead to additional
relativistic energy at last scattering. It should be noted that
the larger expansion rate also increases the helium abundance and hence
these solutions require another variable to offset the effect on helium.

A positive chemical-potential for the electron neutrino can
achieve this offset. In both the decaying tau neutrino and the large chemical
potential scenarios, consistency with all light element abundances can be 
achieved. We, however, consider the decaying tau neutrino scenario less
likely in light of recent results \cite{SuperK} from the
SuperKamiokande collaboration which hint at a mass much less than an MeV
for $\nu_\tau$. To make a comparison between post-BBN entropy change and
non-standard BBN we fix our attention 
on the large chemical-potential scenario. Using $\xi$ to denote the 
ratio between the actual and the BBN baryon density, we have used the
standard BBN code to derive the relation
between $\xi$ and the minimum $\nnu$ required, shown in Fig.~\ref{r_vs_nnu}.
The $\xi$--$\nnu$ relation in the non-standard BBN case is distinct from
that due to post-BBN entropy decrease.

For more than two decades big-bang nucleosynthesis has been both
a critical test of the standard cosmology and a probe of particle
physics (\eg, the limit to the number of light neutrino species) 
and cosmology (\eg, the baryon density).  The cross comparison 
with the baryon density inferred from CMB anisotropy measurements has 
opened a new window for testing the cosmological framework and 
exploring physics beyond the standard model.  Should the current 
discrepancy of about $2\sigma$ persist, the resolution may well 
involve non-standard BBN (neutrino chemical potentials) or new physics
(entropy change due to new particles).  If so, this would be an even 
more impressive achievement for cosmology than the BBN limit 
to the number of neutrino species which itself was ultimately 
confirmed by laboratory experiment. 

\acknowledgements
We acknowledge support by the DOE at Chicago, 
and by the DOE and NASA grant NAG 5-7092 at Fermilab.


\begin{thebibliography}{99}

\bibitem{mather99} J.C. Mather \etal, \apj {\bf 512}, 511 (1999).

\bibitem{tyson} See \eg, M.S. Turner \& J.A. Tyson, Rev. Mod. Phys.
{\bf 71}, S145 (1999).

\bibitem{dns_mst_99} D.N. Schramm \& M.S. Turner, Rev. Mod. Phys.
{\bf 70}, 303 (1998).

\bibitem{tytler_00}D. Tytler \etal, Physica Scripta, {\bf T85}, 12 (2000). 

\bibitem{bntt} S. Burles, K.M. Nollett, J.N. Truran \& M.S.Turner,
\prl {\bf 82}, 4176 (1999); K.M. Nollett \& S. Burles, Phys. Rev. D
{\bf 61}, 123505 (2000).

\bibitem{boom00} P. de Bernardis \etal, Nature {\bf 404}, 955 (2000).

\bibitem{max00} S. Hanany \etal, \apjl, in press (2000) (astro-ph/0005123).

\bibitem{boom_max} A.H. Jaffe \etal, astro-ph/0007333.

\bibitem{other}M. Tegmark \& M. Zaldarriaga, \prl, {\bf 85}, 2240 (2000);
A.E. Lange \etal, astro-ph/0005004; 
W. Hu \etal, astro-ph/006436; 
S. Esposito \etal, astro-ph/0007419;
M. White \etal, astro-ph/0004385.

\bibitem{mst_99} See \eg, M.S. Turner, Physica Scripta {\bf T85}, 210 (2000),
or A. Dekel \etal, in {\em Critical Diaglogues in Cosmology},
edited by N. Turok (World Scientific, Singapore, 1997), p.~175.

\bibitem{H0} J.R. Mould \etal, \apj {\bf 529}, 786 (2000).

\bibitem{os_review} K.A. Olive, G. Steigman \& T.P. Walker,
Phys. Rep. {\bf 333}, 389 (2000).

\bibitem{eta_BBN_note}{The entropy transferred to photons from $e^\pm$
annihilations decreases the baryon-to-photon ratio by 4/11. $\eta$
denotes the post $e^\pm$-annihilation ratio; the deuterium abundance 
depends only on this value.}

\bibitem{kt}E.W. Kolb \& M.S. Turner, {\em The Early Universe},
(Addison-Wesley, Redwood City, CA, 1990), Ch.~5.

\bibitem{bh} J.G. Bartlett \& L.J. Hall, \prl {\bf 66}, 541 (1991).

\bibitem{Hu_review}
W. Hu, N. Sugiyama \& J. Silk, Nature {\bf 386}, 37 (1997).

\bibitem{LS_note} 
{Two implicit assumptions have been made: baryon-number
conservation, and the standard theory of recombination which 
determines $\zls$ precisely.  
The current data support the standard theory of recombination; future 
data will test this assumption more precisely.}

\bibitem{lopezetal} R. Lopez \etal, \prl {\bf 82}, 3952 (1999).

\bibitem{post_bbn_nuc} It is also possible that
post-BBN hadronic and EM showers, which destroy $^4$He and 
produce D, could explain $r < 1$. This does
not change $\nnu$. 
See \eg, S. Dimopoulos \etal, \apj {\bf 330}, 545 (1988). 

\bibitem{tau_decay} G. Gyuk \& M.S. Turner, \prd {\bf 50},
6130 (1994); A.D. Dolgov, S.H. Hansen, S. Pastor \& D.V. Semikoz,
Nucl. Phys. {\bf B548}, 385 (1999).

\bibitem{steigman} H. Kang \& G. Steigman, Nucl. Phys. {\bf B372}, 494
(1992).

\bibitem{resolve} J. Lesgourgues \& M. Peloso, \prd {\bf 62}, 81301 (2000);
S.H. Hansen \& F.L. Villante, Phys. Lett. {\bf B486}, 1 (2000); 
S. Esposito \etal, JHEP {\bf 09}, 038 (2000).

\bibitem{SuperK}S. Fukuda {\etal}, hep-ex/0009001.

\end{thebibliography}
\end{document}